\documentclass[12pt]{article} 
\usepackage{amsmath,amssymb,eucal,bbm} 
\usepackage[all,cmtip]{xy}
\font\tenrm=cmr10

\font\cmssl=cmss10 at 12 pt  
   
\font\bigss=cmssdc10 scaled 2300

\font\cmsslll=cmss10 at 14 pt

\renewcommand{\a}{\alpha}  
\renewcommand{\b}{\beta}

\newcommand{\g}{\gamma}

\renewcommand{\l}{\lambda}  
\renewcommand{\o}{\omega}

\newcommand{\z}{\zeta}

\newcommand{\G}{\Gamma}

\renewcommand{\O}{\Omega}

\newcommand{\bC}{\mathbb{C}}  
\newcommand{\bR}{\mathbb{R}}  
\newcommand{\bZ}{\mathbb{Z}}

\renewcommand{\gg}{\mathfrak{g}}  
  
\newcommand{\gk}{\mathfrak{k}}

\newcommand\Sp{\mathrm{Sp}}  
\newcommand\GL{\mathrm{GL}}  
\newcommand\SL{\mathrm{SL}}  
\newcommand\SO{\mathrm{SO}}  
\newcommand\SU{\mathrm{SU}}  
\newcommand\U{\mathrm{U}}  
  
\newcommand{\id}{{\mathbbm{1}}}   
    
\renewcommand{\square}{\kern1pt\vbox  
               {\hrule height 0.6pt\hbox{\vrule width 0.6pt\hskip 3pt  
    \vbox{\vskip 6pt}\hskip 3pt\vrule width 0.6pt}\hrule height0.6pt}  
    \kern1pt}  

\newcommand{\ra}{\rightarrow}
\newcommand{\lra}{\longrightarrow}

\newcommand{\ol}{\overline} 
 
\newtheorem{Pb}{Problem}

\newtheorem{Th}{Theorem}  
\newtheorem{Prop}{Proposition}  
  
\newtheorem{Cor}{Corollary}  
\newtheorem{Lem}{Lemma}  
\newtheorem{Def}{Definition} 
\newcommand{\bP}{\begin{Pb}\ \ } 
\newcommand{\eP}{\end{Pb}}  
\newcommand{\bt}{\begin{Th}\ \ }  
\newcommand{\et}{\end{Th}}  
\newcommand{\bp}{\begin{Prop}\ \ }  
\newcommand{\ep}{\end{Prop}}  
\newcommand{\bc}{\begin{Cor}\ \ }  
\newcommand{\ec}{\end{Cor}}  
\newcommand{\bl}{\begin{Lem}\ \ }  
\newcommand{\el}{\end{Lem}}  
\newcommand{\bd}{\begin{Def}\ \ }  
\newcommand{\ed}{\end{Def}}  
  
\newcommand{\pf}{\noindent{\it Proof:\ \ }}  
\newcommand{\qed}{\hfill\square}  
\newcommand{\n}{\nabla}

\newcommand{\be}{\begin{equation}}  
\newcommand{\ee}{\end{equation}}  
  
\newcommand\re[1]{(\ref{#1})}  
\newcommand{\arr}{\begin{array}{rlll}}  
\newcommand{\ea}{\end{array}}  
\newcommand{\bea}{\begin{eqnarray}}  
\newcommand{\eea}{\end{eqnarray}}  
\newcommand{\bean}{\begin{eqnarray*}}  
\newcommand{\eean}{\end{eqnarray*}}  
  
\catcode`@=11  
\@addtoreset{equation}{section}  
\catcode`@=12

\begin{document}  
 \rightline{} 
\vskip 1.5 true cm  
\begin{center}  
{\bigss  Completeness in supergravity constructions}\\[.5em]
\vskip 1.0 true cm   
{\cmsslll  V.\ Cort\'es$^1$, T.\ Mohaupt$^2$ and H.\ Xu$^1$} \\[3pt] 
$^1${\tenrm   Department of Mathematics\\  
and Center for Mathematical Physics\\ 
University of Hamburg\\ 
Bundesstra{\ss}e 55, 
D-20146 Hamburg, Germany\\  
cortes@math.uni-hamburg.de}\\[1em]  
$^2${\tenrm Theoretical Physics Division\\ 
Department of Mathematical Sciences\\ 
University of Liverpool\\
Liverpool L69 3BX, UK\\  
Thomas.Mohaupt@liv.ac.uk}\\[1em] 
January 26, 2011 
\end{center}  
\vskip 1.0 true cm  
\baselineskip=18pt  
\begin{abstract}  
\noindent  
We prove that the supergravity r- and c-maps 
preserve completeness. As a consequence, 
any component $\mathcal{H}$ of a hypersurface $\{ h=1\}$ defined by 
a homogeneous cubic polynomial $h$ such that 
$-\partial^2h$ is a complete Riemannian metric on 
$\mathcal{H}$  defines a complete projective special K\"ahler manifold
and any complete projective special K\"ahler manifold 
defines a complete quaternionic K\"ahler manifold of 
negative scalar curvature.  We classify 
all complete quaternionic K\"ahler manifolds of dimension less or equal to 
12 which are obtained in this way and describe some complete examples
 in 16 dimensions.

\end{abstract}

\tableofcontents
\section*{Introduction}
The supergravity r-map and the supergravity c-map are
geometric constructions known to theoretical physicists working
in supergravity and string theory. They can be obtained
by dimensional reduction of the vector multiplet sector of supergravity theories
with eight real supercharges from 5 to 4 and from 4 to 3 spacetime 
dimensions, respectively. The reduction from 4 to 3 dimensions was 
worked out by Ferrara and Sabharwal \cite{FS}, who used the resulting explicit 
local description of the c-map to prove that it maps a projective 
special K\"ahler manifold (see Definition \ref{psKDef}) of real dimension 
$2n$  defined by a holomorphic prepotential $F=F(z^0,\ldots ,z^n)$ 
homogeneous of degree two  
to a quaternionic K\"ahler manifold of dimension $4n+4$ 
of negative scalar curvature. Similarly, it was shown by de Wit and 
Van Proeyen \cite{DV2} that the r-map maps projective special real manifolds   
(see Definition \ref{psrDef}) of dimension $n$ defined by a cubic
polynomial $h=h(x^1,\ldots ,x^n)$  
to projective special K\"ahler manifolds of dimension $2n+2$. 
  
Despite some recent advances in the geometric understanding of the
r-map \cite{AC2} and the c-map \cite{H}, as well as in finding 
new formulations of the c-map within the formalisms of supergravity
\cite{DDKV,DRV,DS,RVV},  
very little is know about global 
geometric properties of these constructions.
In recent approaches to hypermultiplet moduli spaces, there have been 
efforts aiming at the computation of 
quantum corrections to the Ferrara-Sabharwal metric inspired by the 
work of Gaiotto, Moore and Neitzke on wall crossing \cite{GMN}. To our present 
knowledge  this has been successfull only in the limit in which gravity 
decouples, that is in the framework of hyper-K\"ahler geometry.  
The problem of completeness of quaternionic K\"ahler metrics 
obtained as deformations (e.g.\ by quantum corrections) of quaternionic
K\"ahler metrics constructed by the c-map is an interesting subject for
future investigation. 

The main results of this paper are  Theorem \ref{rthm}  
and Theorem \ref{cthm}, which state that the supergravity r-map and the 
supergravity c-map preserve completeness.   
As a consequence, we obtain an effective method for the construction
of complete projective special K\"ahler and quaternionic K\"ahler manifolds
starting from certain real affine hypersurfaces defined by homogeneous cubic 
polynomials $h$. Any such hypersurface of dimension $n$ defines
a quaternionic K\"ahler manifold of dimension $4n+8$, that is 
a point defines an 8-fold,  a curve defines a 12-fold, a surface defines a 
16-fold etc.  The study of the completeness of the quaternionic K\"ahler 
manifold is reduced to that of the completeness of the cubic hypersurface
$\mathcal{H}\subset \bR^{n+1}$ 
with respect to the Riemannian metric $-\partial^2h|_{\mathcal{H}}$.  We 
show how to obtain interesting complete examples and even classification 
results in low dimensions. In particular, we find two complete 
inhomogeneous examples, see Corollary \ref{classCor} b) and Example 3.
The homogeneous examples are automatically complete, as is any 
complete Riemannian manifold. Moreover, all known examples of
homogeneous projective special K\"ahler manifolds with exception of
the complex hyperbolic spaces are in the
image of the r-map and all known examples of homogeneous 
quaternionic K\"ahler manifolds of negative scalar curvature 
with exception of the quaternionic
hyperbolic spaces are in the image of the c-map, see \cite{DV2} and 
references therein. The known examples include the 
homogeneous projective special K\"ahler manifolds of semisimple 
groups classified in \cite{AC1} and the quaternionic 
K\"ahler manifolds admitting a simply transitive (completely) solvable 
group of isometries 
classified in \cite{A,C1}. The first class contains only Hermitian symmetric 
spaces of noncompact type, whereas the second class contains all quaternionic 
K\"ahler symmetric spaces of noncompact type as well as all known 
nonsymmetric homogeneous quaternionic K\"ahler manifolds.  We plan to work towards general completeness results for
hypersurfaces in higher dimensions in the future. 

One of the inhomogeneous complete examples is the
`quantum STU model' \cite{AFT,AFGNT,KV,KKLMV}. 
This model, or actually family of models,
can be constructed by compactification
of the heterotic string on manifolds with holonomy contained
in $SU(2)$ and with instanton numbers 
$(12,12)$ or $(13,11)$ or $(10,4)$. 
The special real, special K\"ahler and special quaternionic 
manifold occur as moduli spaces of compactifications on 
$K3\times S^1$, $K3\times T^2$ and $K3 \times T^3$, respectively.
The qualification  `quantum' refers
to the quantum corrections to the cubic part of the underlying Hesse
potential 
\[
h = STU \rightarrow h = STU + \frac{1}{3}U^3 \;,
\]
which makes the corresponding manifolds inhomogeneous.
This deformation is captured by the triple intersection forms of dual
models, which are compactifications on 
Calabi-Yau threefolds which are elliptic fibrations over the
Hirzeburch surfaces $F_0$, $F_1$, $F_2$, respectively \cite{KV,MV,W}.
The choice of the base of the fibration corresponds to the choice
of instanton numbers in the dual heterotic model. 
The heterotic compactifications on $K3 \times T^2$ are dual 
to type-IIA compactifications on the corresponding Calabi-Yau
threefolds, and the vector
multiplet moduli spaces of these models are the complexified K\"ahler 
cones of the Calabi-Yau threefolds. The type-IIA model
has an `M-theory lift' to a compactification of eleven-dimensional
M-theory on the same Calabi-Yau threefold, which is then 
dual to the heterotic compactification on $K3 \times S^1$. 
The moduli spaces
of five-dimensional vector multiplets correspond to a fixed volume
slice of the (real) K\"ahler cones of the underlying Calabi-Yau 
threefolds. Some properties of the 
special K\"ahler and special real metrics occuring in the vector multiplet
sectors of these models have been discussed in the physics literature
\cite{LSTY,ChouEtAl,MM}. The geodesically complete spaces considered
in this paper are the natural choices of scalar manifolds if these
models are considered as supergravity models. The moduli spaces relevant
for string theory are sub-domains of these manifolds, as discussed 
in \cite{MV,W,LSTY,ChouEtAl,MM}.

In the last section we give a geometric interpretation of the complex  
$(n+1)\times (n+1)$-matrix ${\cal N}=({\cal N}_{IJ})$, which defines the 
nontrivial part of the  Ferrara-Sabharwal metric. 
We show that ${\cal N}$ defines 
a Weil flag which is precisely the image of the Griffiths flag   
associated with the variation of Hodge structure of weight 3 
defined by the underlying affine special K\"ahler manifold under a natural 
$\Sp (\bR^{2n+2})$-equivariant map from Griffiths to Weil flags, 
see Corollary \ref{Cor5}.
Furthermore, ${\cal N}$ is canonically associated with 
a positive definite metric which differs from the (indefinite) affine 
special K\"ahler metric by a canonical sign switch, see Corollary \ref{Cor6}. 
Using this geometric insight, we are able to extend the c-map and 
our completeness result to special K\"ahler manifolds which do 
not admit a global description by a single 
prepotential, see Theorem \ref{globalThm}.   

\section{Completeness of metrics on product manifolds and bundles}  
Let us recall that a Riemannian manifold $(M,g)$ 
is called {\cmssl complete} if it is complete as a metric space, i.e.\
if every Cauchy sequence in $M$ converges. 
The basic result in Riemannian geometry concerning completeness
is the following theorem of Hopf and Rinow, 
cf.\ \cite{O} Ch.\ 5, Thm.\ 21. 
\bt {\bf (Hopf-Rinow)} \label{th:Hopf-Rinow} For a Riemannian 
manifold $(M,g)$ the following conditions are equivalent: 
\begin{enumerate}
	\item[(i)] $M$ is complete. 
	\item[(ii)] $M$ is geodesically complete, i.e.\
every inextendible geodesic ray has infinite length.  
	\item[(iii)] Any closed and bounded subset of $M$ is compact. 
\end{enumerate}
\et

We give another equivalent formulation of completeness 
in terms of (smooth) curves $\g : I \ra M$, where $I\subset \bR$ 
is a (nondegenerate) interval. 

\bl \label{lem:1}A Riemannian manifold $(M,g)$ is complete if and only if 
every curve $\g : I \ra M$  which is not contained 
in any compact subset of $M$ has infinite length. 
\el  

\pf ``$\Rightarrow$'' Let us assume that $M$ is complete and 
that $\g : I \ra M$ is not contained in any compact set. Then
$\g (I)\subset M$ is unbounded, since it is not
contained in any ball. Here we use the fact that closed balls
are compact in any complete Riemannian manifold, by Theorem 
\ref{th:Hopf-Rinow} (iii). 
Clearly, an unbounded curve has infinite length.\\ 
``$\Leftarrow$'' If $M$ is not complete, then there exists
an inextendible geodesic ray $\g : [0,L)\ra M$ of finite
length $L$, by Theorem \ref{th:Hopf-Rinow} (ii). The ray $\g$ is not contained
in any compact set $K$, because otherwise $\g(t)$ would converge
to a limit point in $K$ for $t\ra L$ and $\g$ could then be extended 
to a geodesic ray $\tilde{\g} : [0,\tilde{L})\ra M$ for some 
$\tilde{L}>L$. 
\qed

Let $M=M_1\times M_2$ be a product manifold 
and denote by $\pi_i : M\ra M_i$, $i=1,2$, the 
projections. We consider  
Riemannian metrics $g$ of the form
\begin{equation} g= g_1 + g_2,
\end{equation}
where $g_1$ 
is (the pullback of)  
a Riemannian metric 
on $M_1$ 
and $g_2\in \G (\pi_2^*S^2T^*M_2)$ 
is a family of Riemannian metrics on $M_2$ 
depending on a parameter $p\in M_1$.
Notice that the tensors $g_1$ and $g_2$ take the 
form 
\[ g_1=\sum g_{ab}^{(1)}(x)dx^adx^b,\quad g_2=\sum 
g_{\a \b }^{(2)}(x,y)dy^\a dy^\b, \] 
with respect to local coordinates $x=(x^a)$ on $M_1$ and $y=(y^\a)$ on $M_2$. 

We will assume that the tensor field $g_2$ satisfies the following condition: 
\begin{enumerate}
\item[(C)] For all compact subsets 
$A\subset M_1$ there exists a complete Riemannian metric $g_A$ on 
$M_2$ such that $g_2\ge \pi_2^*g_A$ on $A\times M_2 \subset M$.    
\end{enumerate}
\bl \label{lem:2} 
Assume that $(M_1,g_1)$ is complete and that $g_2$ satisfies the condition 
(C). Then $(M,g)$ is complete.  
\el 

\pf Let $\g = (\g_1,\g_2) : I \ra M=M_1\times M_2$ be a curve 
which is not contained in any compact subset $K\subset M$. 
In view of Lemma \ref{lem:1},  it suffices to show that 
any such curve $\g$ has infinite length. {}From the assumption
on $\g$ it follows that 
\begin{itemize}
\item[(i)]
$\g_1 : I \ra M_1$ is not 
contained in any compact subset $K_1\subset M_1$ or 
\item[(ii)] $\g_2 : I \ra M_2$ is not 
contained in any compact subset $K_2\subset M_2$. 
\end{itemize}
(Otherwise, $\g$ would be contained in a compact set $K=K_1\times K_2$.) 
In case (i), $L(\g_1) = \infty$, by the completeness of $(M_1,g_1)$ and
Lemma \ref{lem:1}.  
Comparing lengths we obtain    
\[ L(\g ) \ge L(\g_1) = \infty  \]
and, hence, $L(\g )=\infty$. 
If (i) is not satisfied, then $\g_1 (I)$ is contained in a 
compact set $A=K_1\subset M_1$. By the completeness of $(M_2,g_A)$ and 
property (ii), we now have that  $L_{g_A}(\g_2) = \infty$. 
Now it suffices to compare the lengths: 
$L(\g ) \ge L_{g_A}(\g_2) = \infty$.
\qed 
\bt \label{thm:main} 
Let $(M_1,g_1)$ be a complete Riemannian manifold and
$(g_2(p))_p$ a smooth family of $G$-invariant Riemannian
metrics on a homogeneous manifold $M_2=G/K$, depending on a parameter
$p\in M_1$. Then the Riemannian metric $g=g_1+g_2$ on
$M=M_1\times M_2$ is complete. Moreover, the action of $G$ on $M_2$
induces an isometric action of $G$ on $(M,g)$. 
\et   

\pf The last assertion is obvious. For the completeness of $(M,g)$, 
it suffices to check that $g_2$ satisfies the 
condition (C) of Lemma \ref{lem:2}. We use the natural one-to-one
correspondence between  $G$-invariant Riemannian metrics 
on $M_2=G/K$ and $K$-invariant scalar products on the vector
space $T_oM_2\cong \gg/\gk$, $o=eK$. Under this 
correspondence, the family $(g_2(p))_p$ corresponds to
a family $(\beta (p))_p$ of scalar products. For every compact subset 
$A\subset M_1$ the family $(\beta (p))_{p\in A}$ is uniformly
bounded from below by a scalar product $\b_A$:
\[ \b (p) \ge \b_A,\quad\mbox{for all}\quad p\in A.\] 
This implies 
\[ g_2(p)\ge g_A,\quad\mbox{for all}\quad p\in A,\]  
for the $G$-invariant Riemannian
metric $g_A$ associated with $\b_A$. Now it
suffices to remark that $g_A$ is complete, as is any 
$G$-invariant Riemannian metric on $G/K$. 
\qed 
\bc \label{cor:1} 
Let $g_U=\sum g_{ab}dx^adx^b$ be a complete Riemannian
metric on a domain $U\subset \bR^n$. Then the metric 
\begin{equation} \label{e:rmap} g_M=\frac{3}{4}\sum g_{ab}(x)(dx^adx^b+ dy^ady^b)
\end{equation}
on $M=U\times \bR^n$ is complete. The action of $\bR^n$ by translations
in the $y$-coordinates is isometric and the projection 
$M\ra U$ is a principal fiber bundle with structure group $\bR^n$. 
The submanifold $U= U\times \{ 0\} \subset U\times \bR^n=M$ is totally
geodesic.  
\ec
(The factor $\frac{3}{4}$ is introduced only in order to obtain
the usual normalization of the projective special K\"ahler metric  
for the r-map defined in the next section.) 

\pf For the completeness it suffices to apply Theorem \ref{thm:main} to the case
$M_2=G=\bR^n$, $g_2=\sum g_{ab}(x)dy^ady^b$. $U\subset M$ is totally 
geodesic as the fixed point set of the isometric
involution $(x,y)\mapsto (x,-y)$.\qed 
\subsection{Generalisation to the case of bundles }{\label{bundlesSec}}
More generally, for later applications we consider now a bundle
$\pi : M \ra M_1$ with standard fiber $M_2$ 
over a  Riemannian manifold
$(M_1,g_1)$. We suppose that the total space $M$ 
is endowed with a Riemannian metric $g$ such that
for all $p\in M_1$ there exists a neighbourhood $U\subset 
M_1$ and a local trivialisation $\pi^{-1}(U) \cong U\times M_2$
with respect to which the metric takes the form
\be \label{g2} g|_{\pi^{-1}(U)} = g_1|_U + g_2^U,\ee
where $g_2^U$ is a smooth family of Riemannian metrics
on $M_2$ depending on a parameter in $U$. Such metrics $g$ will be
called {\cmssl bundle metrics}. We will assume that $g_2^U$ 
satisfies the condition (C) for all compact subsets $A\subset U$.  
Lemma \ref{lem:2} and Theorem \ref{thm:main} have the following 
straightforward generalisations:
\bl Assume that $(M_1,g_1)$ is complete and that the local fiber metrics  
$g_2^U$ in equation \re{g2} satisfy the condition (C) 
for all compact subsets $A\subset U$. Then the bundle metric $g$
is complete. 
\el  
\bt \label{auxThm} Let $g$ be a bundle metric on a bundle $\pi : M \ra M_1$ 
over a complete Riemannian manifold
$(M_1,g_1)$ and assume that the standard fiber is a homogeneous space  
$M_2=G/K$ and that the fiber metrics  
$g_2^U$ are $G$-invariant. Then $(M,g)$ 
is complete. 
\et 
\section{The generalized  supergravity r-map preserves completeness}
Let $U\subset \bR^n$ be a domain which is invariant under multiplication 
by positive numbers and let $h:U\ra \bR^{>0}$ be a smooth function which is  
homogeneous of degree $d\in \bR \setminus\{ 0,1\}$. Then 
\[ \mathcal{H} :=\{ x\in U| h(x)=1\}\subset U\]
is a smooth hypersurface and $U= \bR^{>0}\cdot \mathcal{H}
\cong \bR^{>0}\times \mathcal{H}$. 
We will assume that $-\frac{1}{d}\partial^2h$ is positive  
definite on $T\mathcal{H}$. This easily implies that 
$-\frac{1}{d}\partial^2h$ is a 
Lorentzian (if $d>1$) or Riemannian (if $d<1$) 
metric on $U$ which restricts to 
a Riemannian metric 
$g_{\mathcal{H}}$ on $\mathcal{H}$. 
\bd \label{psrDef} The Riemannian manifold $(\mathcal{H}, g_{\mathcal{H}})$
is called a {\cmssl projective special real manifold} if, in addition, $h$ 
is a polynomial function of degree $d=3$. 
\ed 

\noindent
{\bf Example} Let $V=H^{1,1}(X,\bR )$ be the $(1,1)$-cohomology of a compact 
K\"ahler manifold $X$ of complex dimension $d\ge 2$ and $U\subset V$ the 
K\"ahler cone. We define a homogeneous polynomial $h$ of degree $d$ on $V$
by 
\[ h(a) = a^{\cup d}=a\cup \cdots \cup a,\quad a\in V.\] 
The polynomial $h$ defines a positive function on $U$ and the 
metric $g_\mathcal{H}$ defined above is positive definite on 
the hypersurface
\[ \mathcal{H} := \{ x\in U| h(x)=1\}\subset V\cong \bR^n,\quad n=h^{1,1}(X).\]
This follows from the Hodge-Riemann bilinear relations, see \cite{We} Chap.\ V, 
Sec.\ 6, which 
imply that $g_\mathcal{H}$ is positive definite on the primitive 
cohomology $H^{1,1}_0(X,\bR )=T_\kappa\mathcal{H}$ defined as 
the kernel of the cup product with 
$\kappa^{d-1}: H^2(X,\bR ) \ra H^{2d}(X,\bR )$, where $\kappa\in U$ is a  
K\"ahler class. If $d=3$ then ($\mathcal{H}$, $g_\mathcal{H}$) 
is a projective special real manifold. 


We endow $U$ with the Riemannian metric
\begin{equation} \label{e:gU} 
g_U = -\frac{1}{d}\partial^2 \ln h \end{equation} 
and $M=TU\cong U\times \bR^n$ with the metric \re{e:rmap}. 
\bp \label{prop:product} $(U,g_U)$ is isometric to the product
$(\bR\times \mathcal{H},dr^2+g_{\mathcal{H}})$. 
\ep
  
\pf This is a straightforward calculation using
the diffeomorphism
\[ \bR\times \mathcal{H}\ra U,\quad (r,x) \mapsto e^rx,\]
and the formula  \re{e:gU}.  
\qed 

\bd 
The correspondence $(\mathcal{H},g_{\mathcal{H}})\mapsto (M,g_M)$ is called 
the {\cmssl generalized supergravity r-map}. The restriction 
to polynomial functions $h$ of degree $d=3$ is called the 
{\cmssl supergravity r-map}.  
\ed
Next we need to recall the notion of a projective special K\"ahler manifold 
\cite{ACD,DV1,F}. It is best explained starting from the notion of 
a conical special K\"ahler manifold, cf.\ \cite{ACD,CM}. For the 
purpose of this paper, we shall restrict the signature of the metric
by the condition (iv) in the following definition.  
 
\bd \label{conDef} 
A {\cmssl conical special K\"ahler manifold} $(M,J,g,\n ,\xi )$ is a 
pseudo-K\"ahler manifold $(M,J,g)$ endowed with a flat torsionfree 
connection $\n$ and a vector field $\xi$ such that 
\begin{enumerate}
\item[(i)] $\n \o=0$, where $\o = g (\cdot , J\cdot )$ is the K\"ahler form,
\item[(ii)] $d^\n J=0$, where $J$ is considered as a one-form with values
in the tangent bundle, 
\item[(iii)] $\n \xi = D\xi = {\rm Id}$, where $D$ is the Levi-Civita
connection,    
\item[(iv)] $g$ is positive definite on the plane $\mathcal{D} = \mathrm{span}
\{ \xi , J\xi \}$ and negative definite on $\mathcal{D}^\perp$.   
\end{enumerate}
\ed
It is shown in \cite{ACD,CM} that the geometric data of a conical special 
K\"ahler manifold  can be locally encoded in a holomorphic function $F$ 
homogeneous of degree 2 defined in a domain $M_F\subset \bC^n$, 
see Theorem 2 and Proposition 6 \cite{CM}.  In fact, 
$F$ is the generating function of 
a holomorphic Lagrangian immersion  $M_F\ra \bC^{2n}$, which induces
on $M_F$ the structure of a conical special K\"ahler manifold. 
$F$ is called the  {\cmssl holomorphic prepotential} in the supergravity 
literature \cite{DV1}. Under the assumptions of Definition \ref{conDef}, 
the 
vector fields $\xi$ and $J\xi$ define a holomorphic action 
of a two-dimensional Abelian Lie algebra. We will assume that 
this infinitesimal action lifts to a 
principal $\bC^*$-action on $M$ with the base manifold $\bar{M}=M/\bC^*$.
Then the hypersurface $S:= \{ p\in M| g(\xi (p),\xi (p))=1\} \subset M$
is an $S^1$-principal bundle over $\bar{M}$. The principal action is 
isometric,  
since it is generated by the Killing vector field $J\xi$. Therefore, 
the Lorentzian metric $g_S= -g|_S$ induces a Riemannian metric $\bar{g}$ 
on $\bar{M}$, which is easily seen to be K\"ahlerian. In fact, the negative 
definite K\"ahler manifold $(\bar{M},-\bar{g})$ 
is precisely the K\"ahler reduction of $(M,g)$ 
with respect to the above isometric 
Hamiltonian $S^1$-action for a positive level of the 
moment map, which is $\mu = \frac{g(\xi ,\xi )}{2}$. 
\bd \label{psKDef} The K\"ahler manifold $(\bar{M},\bar{g})$  
is called a {\cmssl projective special K\"ahler manifold}.  
The metric $\bar{g}$ is called a {\cmssl projective special K\"ahler metric}. 
\ed 
A standard example of a projective special K\"ahler metric is the 
Weil-Petersson metric on the  space of 
complex structure deformations of a Calabi-Yau 3-fold, 
that is of a compact K\"ahler manifold with holonomy 
$\SU (3)$, see, for instance, \cite{T,C2}.  
\bt \label{rthm} 
The generalized supergravity r-map maps complete Riemannian manifolds 
$(\mathcal{H},g_{\mathcal{H}})$ as above to complete K\"ahler manifolds $(M,g_M)$
with a free isometric action of the vector group $\bR^n$. 
The supergravity r-map maps complete projective special real manifolds 
to complete projective special K\"ahler manifolds. 
\et

\pf The isometric action of $\bR^n$ exists for general metrics as in 
Corollary \ref{cor:1}.  The K\"ahler property follows from the fact that 
the metrics $g_U$ considered here are of Hessian type, cf.\ 
\cite{AC2} Prop.\ 3. 
In fact, $(\mathcal{H},g_{\mathcal{H}})$ is mapped to $M=U\times \bR^n \cong 
\bR^n+iU\subset \bC^n$ with the metric $g_M$ defined by the 
K\"ahler potential $-\ln h(x)$, 
where $(x,y)\in U\times \bR^n=M$ is identified with $\zeta = y+ix
\in\bR^n+iU\subset \bC^n$. If $h$ is a cubic polynomial then a simple
calculation shows that the metric $g_M$ can be also obtained from 
the K\"ahler potential $K(1,\zeta)$ defined by   
\[ K(z)=-\ln 
\left( i\sum _{I=0}^n\left( z^I\bar{F}_I-F_I\bar{z}^I\right) \right),\]
where 
\be \label{prepotEqu} F(z^0,\ldots ,z^n)= h(z^1,\ldots ,z^n)/z^0\ee
is a holomorphic function homogeneous of degree $2$ on the domain 
\[ \tilde{M}:=\{ z^0\cdot (1,\zeta )| z^0\in \bC^*,\quad \zeta \in  \bR^n+iU\}\subset 
\bC^{n+1}.\] 
(It suffices to check that $\frac{i}{|z^0|^2}\sum (z^I\bar{F}_I-F_I\bar{z}^I) 
= 8h(x)$.) 
This shows that $(M,g_M)$ 
is a projective special K\"ahler manifold with
the holomorphic prepotential $F$. The corresponding conical 
special K\"ahler manifold is the $\bC^*$-bundle $\tilde{M}\ra M$
endowed with the affine special K\"ahler metric 
$g_{\tilde{M}}=2\sum_{I,J=0}^n (\mathrm{Im} F_{IJ})dz^Id\bar{z}^J$, 
which has signature $(2,2n)$. 
By Proposition \ref{prop:product} and Corollary \ref{cor:1}, 
the completeness of $(\mathcal{H},g_{\mathcal{H}})$ implies that of $(U,g_U)$ 
and the completeness of $(U,g_U)$ implies that of $(M,g_M)$. 
\qed 
\bc Any complete projective special real manifold $(\mathcal{H},g_{\mathcal{H}})$
of dimension $n-1\ge 0$ admits a canonical realisation as a totally
geodesic submanifold of a complete projective special 
K\"ahler manifold $(M,g_M)$ with a free isometric
action of the group $\bR^n$.  Each orbit of $\bR^n$ is flat
and and intersects
the submanifold $\mathcal{H}\subset M$ orthogonally in exactly one point. 
\ec

\pf The orbits are flat since the metric is translation invariant. 
The hypersurface $\mathcal{H}\subset U \cong \bR^{>0} \times \mathcal{H}$ 
is totally geodesic in virtue of Proposition \ref{prop:product}
and $U\subset M$ is totally geodesic in virtue of Corollary \ref{cor:1}.
Therefore $\mathcal{H}\subset M$ is totally geodesic. 
\qed

Notice that the above proof shows that $\mathcal{H}\subset M=r(\mathcal{H})$ 
is totally geodesic also for the generalized r-map. 
\section{The supergravity c-map preserves completeness}{\label{ccomplsubsec}}
A projective special K\"ahler manifold $(M,g_M)$ which is globally defined 
by a single holomorphic prepotential $F$ is called a {\cmssl projective 
special K\"ahler domain}. Notice that the manifolds in the image of the 
r-map are defined by the prepotential \re{prepotEqu} and, hence, 
are examples of projective special K\"ahler domains. 
Recall \cite{FS} that the {\cmssl supergravity c-map} maps 
projective special K\"ahler domains $(M,g_M)$ of dimension $2n$ to  
quaternionic K\"ahler manifolds $(N,g_N)$ of dimension $4n+4$ and of 
negative scalar curvature.  More generally, we may consider the situation
when the projective special K\"ahler manifold 
$(M,g_M)$ is covered by a collection of charts $U_\a$ on which 
the special K\"ahler geometry is encoded by a prepotential $F_\a$. 
The investigation of this case is postponed to section \ref{lastSec}. 
For a projective special K\"ahler domain $(M,g_M)$, 
the quaternionic K\"ahler metric 
$g_N$ on $N=M\times \bR^{2n+3}\times \bR^{>0}\cong M\times \bR^{2n+4}$ 
is of the form 
\begin{eqnarray} \label{e:2} g_N &=& g_M + g_G,\\\nonumber 
g_G&=& \frac{1}{4\phi^2}d\phi^2 + \frac{1}{4\phi^2}(d\tilde{\phi}
+ \sum (\zeta^Id\tilde{\zeta}_I-\tilde{\zeta}_Id\zeta^I) )^2 
+\frac{1}{2\phi}\sum \mathcal{I}_{IJ}(p) d\zeta^Id\zeta^J\\ 
&&+ \frac{1}{2\phi}\sum \mathcal{I}^{IJ}(p)(d\tilde{\zeta}_I + 
\mathcal{R}_{IK}(p)d\zeta^K)(d\tilde{\zeta}_J + 
\mathcal{R}_{JL}(p)d\zeta^L),
\end{eqnarray}
where $(\tilde{\z}_I, \z^I, \tilde{\phi}, \phi)$, $I=0,1,\ldots, n$, 
are coordinates on $\bR^{2n+4}\supset \bR^{2n+3}\times \bR^{>0}$ 
and the metric is defined for 
$\phi>0$. The matrices $(\mathcal{I}_{IJ}(p))$ and $(\mathcal{R}_{IJ}(p))$
depend only on $p\in M$ and $(\mathcal{I}_{IJ}(p))$ is invertible with
the inverse $(\mathcal{I}^{IJ}(p))$. More precisely,
 \begin{equation} 
\label{FRI}
{\cal N}_{IJ} := 
\mathcal{R}_{IJ} + i\mathcal{I}_{IJ} := 
\bar{F}_{IJ} + 
i \frac{\sum_K N_{IK}z^K\sum_L N_{JL}z^L}{\sum_{IJ}N_{IJ}z^Iz^J} ,\quad 
N_{IJ} := 2 \mathrm{Im} F_{IJ} ,\end{equation}
where $F$ is the holomorphic prepotential with respect
to some system of special holomorphic coordinates $z^I$ on the 
underlying conical special K\"ahler manifold $\tilde{M}\ra M$. 
Notice that the expressions are homogeneous of degree zero and, hence, 
well defined local functions on $M$. Also note that our conventions are
slightly different from those in \cite{FS}. The prepotentials are related
by $F=\frac{i}{4} F^{[FS]}$, while $N=N^{[FS]}$. Also note that 
$\mathcal{I} = - \mathcal{R}^{[FS]}$, and therefore $\mathcal{I}$ 
is positive definite. 

Let $(M,g_M)$ be a special K\"ahler domain and $N=M\times G$ the corresponding
quaternionic K\"ahler manifold with the Ferrara-Sabharwal 
metric $g_N=g_M+g_G$.   We will show that, for fixed $p\in M$,  
$g_G(p)$ can be considered as a left-invariant Riemannian metric 
on a certain Lie group diffeomorphic to $\bR^{2n+4}$.  We define the Lie group
$G$ by putting the following group multiplication on $\bR^{2n+4}$: 
\[ (\tilde{\z},\, \z,\, \tilde{\phi},\, \phi )\cdot 
(\tilde{\z}',\, {\z'},\,  \tilde{\phi}',\, \phi' ) := 
(\tilde{\z} + e^{\phi/2}\tilde{\z}',\, \z +e^{\phi/2}\z',\, 
\tilde{\phi} + e^\phi\tilde{\phi}' + e^{\phi/2}(\tilde{\z}^t\z'-
{\z'}^t\tilde{\z}),\, \phi + \phi' ), 
  \]
where $\z^t=(\z^0,\ldots,\z^n)$ is the transposed of the
column vector $\z$. We remark that 
$G$ is isomorphic to the solvable Iwasawa subgroup of $\SU (1,n+2)$,
which acts simply transitively on the complex hyperbolic space
of complex dimension $n+2$. $G$ is a rank one solvable extension
of the $(2n+3)$-dimensional Heisenberg group. We can realise 
it as a group of affine transformations of  $\bR^{2n+4}$ by associating
to an element $(\tilde{v} , v ,  \a , \l )\in G=\bR^{2n+4}$ the affine
transformation
\begin{equation} \label{e:isom} 
(\tilde{\z},\, \z ,\,  \tilde{\phi } ,\, \phi ) \mapsto 
(e^{\l/2}\tilde{\z}+\tilde{v},\, e^{\l/2}\z + v,\,  
e^{\l/2}(\tilde{v}^t\z -v^t\tilde{\z}) + e^\l\tilde{\phi} +\a,\, e^\l \phi
 ).\end{equation}
In virtue of this action, we can identify $G$ with the orbit
of the point $(0,0,1,0)$, which is $L=\bR^{2n+3}\times \bR^{>0}
\subset \bR^{2n+4}$. This identification is simply 
\[ G\ni (\tilde{v}, v , \a ,  \l )\mapsto (\tilde{v}, v,  \a , e^\l)\in L.\] 
One can easily check that the following pointwise 
linearly independent  one-forms are  invariant under the 
transformations \re{e:isom} and, hence, define a left-invariant coframe on 
$G\cong L$: 
\begin{eqnarray} 
\eta^I &:=& \sqrt{\frac{2}{\phi}}d\z^I,\quad  
\xi_I\; :=\; \sqrt{\frac{2}{\phi}}\left( d\tilde{\z}_I +
\sum \mathcal{R}_{IK}(p)d\z^K\right),\\ \label{coframeEqu}
\xi_{n+1} &:=&\frac{d\phi}{\phi},\quad
 \eta^{n+1}\; :=\; \frac{1}{\phi}\left(d\tilde{\phi} + 
\sum (\z^Id\tilde{\z}_I -\tilde{\z}_Id\z^I)\right). \nonumber
\end{eqnarray}
This shows, in particular,  
that the metric $g_G(p)$ is left-invariant for all
$p\in M$. 

\bt \label{cthm} 
The supergravity c-map maps any complete projective special 
K\"ahler domain 
$(M,g_M)$ to a complete quaternionic K\"ahler manifold $(N,g_N)$ 
of negative scalar curvature, which admits the free isometric
action \re{e:isom} of the group $G$. 
\et

\pf The above description of the c-map, starting from \re{e:2},  shows that 
the  quaternionic K\"ahler manifold $(N,g_N)$ is of the form
$N=M\times G$, $g_N=g_M + g_G$, where 
$g_G$ is a smooth family of left-invariant metrics on the group
$G$ depending on a parameter $p\in M$.  Therefore, Theorem \ref{thm:main} 
shows that $(N,g_N)$ is complete if $(M,g_M)$ is complete. 
\qed 
\bc Any complete projective special 
K\"ahler domain 
$(M,g_M)$ admits a canonical realisation as a totally
geodesic K\"ahler submanifold of a complete quaternionic K\"ahler 
manifold $(N,g_N)$ with a free isometric
action of the group $G$. Each orbit of $G$ is isometric to 
a complex hyperbolic space of holomorphic sectional 
curvature $-4$ and intersects
the submanifold $M\subset N$ orthogonally in exactly one point.  
\ec 
\pf The submanifold $M=M\times \{ e\}\subset M\times G=N$ of the 
quaternionic K\"ahler manifold $N$ defined by the supergravity c-map is the 
fixed point set of the isometric involution 
\[ (\tilde{\z}_I, \z^I,  \tilde{\phi}, \phi) \mapsto (-\tilde{\z}_I, -\z^I,   
-\tilde{\phi}, \phi^{-1}).\]
This implies that $M$ is totally geodesic.  Next we compare $g_G(p)$ 
with the standard left-invariant K\"ahler metric of constant
holomorphic sectional curvature $-1$ on $G$, which originates 
from the simply transitive action of $G$ on the 
complex hyperbolic space. By a linear change of
the coordinates $\zeta^I$ we may assume that the positive
definite symmetric matrix $\mathcal{I}_{IJ}(p)=\delta_{IJ}$. 
Then
\[ g_G(p) = \frac{1}{4}\sum_{i=0}^{n+1}(\xi_i^2 + (\eta^i)^2),\]
where the left-invariant coframe \re{coframeEqu} has the following 
differentials 
\[ d\xi_{n+1} =0,\quad d\eta^{n+1}=-\sum_{I=0}^{n+1}\xi_I\wedge\eta^I,\quad
d\xi_I= -\frac{1}{2}\xi_{n+1}\wedge \xi_I,\quad 
d\eta^I=-\frac{1}{2}\xi_{n+1}\wedge
\eta^I.\]
This shows that, up to multiplication with the factor $1/2$, 
the above coframe is dual to the standard 
orthonormal basis of the elementary K\"ahlerian 
Lie algebra $\mathfrak{g}$ and, hence, 
that $g_G(p)$ is a K\"ahler metric of constant holomorphic 
sectional curvature $-4$. 
\qed

\section{Examples: Complete quaternionic K\"ahler manifolds 
associated to
cubic polynomials}

As an immediate corollary of Theorem \ref{rthm} and Theorem \ref{cthm} we have:
\bt To any complete projective special real manifold 
$(\mathcal{H},g_{\mathcal{H}})$ the composition of the r-map with the c-map
associates a complete quaternionic K\"ahler manifold $(N,g_N)$ 
of negative scalar curvature.
\et  

\noindent
{\bf Example} We can consider the point $\mathcal{H}= \{ 1\} \subset \{ x^3=1\} \subset \bR$ as an example of a projective special real manifold. The 
corresponding complete quaternionic K\"ahler eightfold obtained
by the above construction is the symmetric space $\mathrm{G}_2^*/\SO (4)$
of noncompact type.    
 
A homogeneous cubic polynomial $h\in S^3(\bR^n)^*$ will be called  
{\cmssl hyperbolic (respectively, elliptic)} if there exists a point 
$p\in \{ h=1\} 
:= \{ x\in \bR^n| h(x)=1\}$ such that $\partial^2h$ is negative 
(respectively, positive) definite 
on the tangent space of the hypersurface 
$\{ h=1\}$ at $p$. Such points will be called {\cmssl hyperbolic 
(respectively, elliptic)}. 
Let us denote by $\mathcal{H} = \mathcal{H}(h)\subset \{ h=1\}$ the open subset
of hyperbolic points. It is a projective special real
manifold with the metric $g_{\mathcal{H}}$ given by the
restriction of $-\frac{1}{3}\partial^2h$. 

The classification of complete projective special real manifolds reduces 
to the following two problems:
\bP \label{1stPb} 
Classify all hyperbolic homogeneous 
cubic polynomials up to linear transformations.
In other words, describe the orbit space $S^3(\bR^n)_{hyp}^*/\GL (n)$, where
$S^3(\bR^n)_{hyp}^*\subset S^3(\bR^n)^*$ stands for the open subset of hyperbolic
polynomials. 
\eP

\bP For each hyperbolic homogeneous 
cubic polynomial $h$, determine the components of 
the hypersurface $\mathcal{H}(h)$ which are complete and classify them up to
linear transformations.  
\eP

We will solve these problems in the simplest case, that is for $n=2$.  
This gives the classification of complete projective special real curves.
The classification of complete projective special real surfaces is open, 
but we will give some examples of such surfaces.  

\subsection{Classification of complete cubic curves and corresponding 
12-dimensional quaternionic K\"ahler manifolds}
\bt The orbit space $S^3(\bR^2)_{hyp}^*/\GL (2)$ consists of three
points, which are represented by the polynomials
\[ x^2y,\quad x(x^2-y^2)\quad\mbox{and}\quad x(x^2+y^2).\] 
\et

\pf Let $h\in S^3(\bR^2)_{hyp}^*$. 
Interchanging the variables $x$ and $y$, if necessary, we can assume that
$\deg_x h$, the degree of $h=h(x,y)$ in the variable $x$, is $2$ or $3$. 
Since the Hessian of $h$ is nondegenerate we also have $\deg_y h\ge 1$.\\
Case 1) If $\deg_x h = 3$, then the polynomial $f(x) := h(x,1)$ has degree $3$. 
Any polynomial $f(x)$ of degree $3$ can be brough 
to one of the following forms by an affine transformation in the variable $x$:
\[x^3,\quad cx^2(x-1),\quad c(x+a)x(x-1),\quad  cx(x^2+1),\quad c(x+a)(x^2+1),\]
where $c\neq 0$ and $a>0$ are real constants. The first form is excluded, since
$\deg_y h\ge 1$. This implies that $h$ can be brought to one of the 
following forms by a linear transformation:
\[ x^2(x-y),\quad (x+ay)x(x-y),\quad  x(x^2+y^2),\quad (x+ay)(x^2+y^2),
\quad a>0.\] 
The first polynomial is linearly equivalent to $x^2y$. 
The zero set of the second polynomial in the real projective line
consists of the points $-a, 0,  1\in \bR\subset \bR P^1=P(\bR^2)$.
Since any three pairwise distinct points in the projective line
are related by an element of $\GL (2)$, we can assume that $a=1$.
Finally, the last polynomial can be brought to the form
$x(x^2+y^2)$ by a linear conformal transformation. Therefore,
we are left with the following $3$ normal forms:
\be \label{nfEqu} 
x^2y,\quad x(x^2-y^2)\quad\mbox{and}\quad  x(x^2+y^2).\ee 
Case 2) If $\deg_x h = 2$, then the quadratic polynomial $f(x)=h(x,1)$ can be brought
to one of the following forms by an affine transformation:
\[ \pm x^2,\quad cx(x-1),\quad c(x^2+1),\quad c\neq 0.\]
Therefore $h$ can be brought to one of the following forms by a linear 
transformation: 
\[ x^2y,\quad xy(x-y),\quad (x^2+y^2)y.\]
The last two polynomials are equivalent to the polynomials $x(x^2\mp y^2)$
already included in our list \re{nfEqu}.
  
It remains to check that all $3$ polynomials are indeed hyperbolic.
For dimensional reasons ($n=2$), this is equivalent to
the existence of a point $p=(x,y)\in \bR^2$ such that $h(p)>0$ and 
$D(p):= \det \partial^2h(p) <0$.
\begin{enumerate}
\item[a)] For $h=x^2y$ we have $D=-4x^2$. 
Therefore all points of the curve $\{ h=1\}$ are hyperbolic. 
\item[b)] The same is true for $h=x(x^2-y^2)$, since $D=-4(3x^2+y^2)$.   
\item[c)] For $h=x(x^2+y^2)$ we find $D=4(3x^2-y^2)$. 
The point $\frac{1}{\sqrt[3]{5}}(1,2)\in \{ h=1\}$ is hyperbolic, whereas
$\frac{1}{\sqrt[3]{2}}(1,1)\in \{ h=1\}$ is elliptic.
\end{enumerate}
\qed 

Next we investigate the components of $\mathcal{H}(h)$. 

\bt\begin{enumerate}
\item[a)]  The curve $\{ x^2y=1\}$ consists of hyperbolic points and 
has two equivalent components. 
They are homogeneous and, hence, complete projective special real curves.   
\item[b)] The curve $\{ x(x^2-y^2)=1\}$ consists of hyperbolic points and 
has three equivalent components, which are inhomogeneous 
complete projective special real curves.    
\item[c)] Let $h=x(x^2+y^2)$. The curve 
$\mathcal{H}(h)=\{p\in \bR^2| h(p)=1, D(p)<0\}$, which consists of 
the hyperbolic points of $\{ h=1\}$ has two equivalent
components. They are incomplete. The  curve 
$\{p\in \bR^2| h(p)=1, D(p)>0\}$ which consists of the  
elliptic points of $\{ h=1\}$ is connected and incomplete. 
\end{enumerate}
\et 
   
\pf a) $h=x^2y$. The reflection $x\mapsto -x$ interchanges the two components 
of the curve $\{ h=1\}$ and the subgroup 
$\{ \mathrm{diag}(\l ,\l^{-2})| \l >0\}\subset\GL (2)$ acts transitively 
on each component.\\  
b) $h=x(x^2-y^2)$. The transformation
\[\left( \begin{array}{lr} 
-1/2&1/2\\
-3/2&-1/2\end{array} \right) \in \SL (2)\]
generates a cyclic group of order three, which interchanges
the three components of the curve $\{ h=1\}$. Let us consider
the component 
\[ C := \{ p\in \bR^2| h(p)=1, x>0\}. \]
It is symmetric with respect to the $x$-axis, intersects the
$x$-axis at $x=1$ and approaches the asymptotic lines $y=\pm x$ when
$x\ra \infty$. To prove the completeness of $C$, it suffices to show 
that the length of the upper 
half $C_+ = C\cap \{ y>0\}$ of the curve is infinite. 
A straighforward calculation shows that the 
metric $ds^2=g=-\frac{1}{3}\partial^2h|_C$ is given by
the formula 
\[ \frac{3}{2} g = -3xdx^2+xdy^2+2ydxdy=\frac{3(4x^3-1)}{4x^2(x^3-1)}dx^2.\]
This yields the following asymptotics    
\begin{eqnarray*} \frac{g}{dx^2} &=& \frac{2}{x^2} + O\left( \frac{1}{x^5}
\right)\\
\frac{ds}{dx} &=& \frac{\sqrt{2}}{x} + O\left(\frac{1}{x^4}\right), 
\end{eqnarray*}
which implies that 
the arc length $\int_{1}^x ds= \sqrt{2}\ln x + O(\frac{1}{x^3})$ grows logarithmically with $x$. This shows that $C_+$ has infinite length with 
respect to $g$. A simple calculation shows that the 
automorphism group of $C$ is trivial.\\
c) $h=x(x^2+y^2)$. The two components of $\mathcal{H}(h)=\{ h=1, x< \frac{1}{\sqrt[3]{4}}\}$ 
are interchanged by the
reflection $y\mapsto -y$. They are incomplete due to the  
points of inflections at the boundary points $(\frac{1}{\sqrt[3]{4}},
\pm \frac{\sqrt{3}}{\sqrt[3]{4}})$. 
The same
is true for the curve $\{p\in \bR^2| h(p)=1, D(p)>0\}=\{ h=1, x> 
\frac{1}{\sqrt[3]{4}}\}$. 
\qed   

\bc \label{classCor} 
There exists precisely two complete projective special real curves, up
to linear equivalence:
\begin{enumerate}
\item[a)] $\{ (x,y)\in \bR^2| x^2y=1, x>0\}$ and 
\item[b)] $\{ (x,y)\in \bR^2|x(x^2-y^2)=1,x>0\}$. 
\end{enumerate}
Under the composition of the r- and c-map both give rise to complete 
quaternionic K\"ahler manifolds of dimension 12. In the first case
we obtain the symmetric quaternionic K\"ahler manifold 
\[ \frac{\SO_0(4,3)}{\SO(4)\times \SO(3)}.\]
The second example gives rise to a new complete
quaternionic K\"ahler manifold. 
\ec 
\subsection{Examples of complete cubic surfaces  and corresponding 
16-dimensional  quaternionic K\"ahler manifolds}
{\bf Example 1 (STU model)} The surface $\mathcal{H}=\{ xyz=1, x>0, y>0\} \subset \bR^3$
is a homogeneous projective special real manifold. In fact, the group
$\bR^{>0}\times \bR^{>0}$ acts simply transitively on $\mathcal{H}$ by 
unimodular diagonal matrices. The corresponding quaternionic K\"ahler 
manifold is the symmetric space 
\[ \frac{\SO_0(4,4)}{\SO(4)\times \SO(4)}.\]
 
\noindent 
{\bf Example 2} The surface $\mathcal{H}=\{ x(xy-z^2)=1, x>0 \} 
\subset \bR^3$ is another homogeneous projective special real manifold. 
It admits 
the following simply transitive solvable group of linear automorphisms:  
$(x,y,z)\mapsto (\l^{-2}x, \l^4(\mu x + y + 2\mu z), \l (\mu x + z))$, $\l >0$, 
$\mu\in \bR$. 
The corresponding quaternionic K\"ahler 
manifold is the nonsymmetric homogeneous manifold $\mathcal{T}(1)$ described in 
\cite{C1}.

\noindent 
{\bf Example 3 (quantum STU model)} The surface $\mathcal{H}=\{ x(yz + x^2)=1, 
x<0, y>0\}$
is an inhomogeneous complete projective special real manifold. 
Its automorphism 
group is one-dimensional. The maximal connected subgroup is given by: 
$(x,y,z)\mapsto (x,\l y, \l^{-1}z)$, $\l >0$. 
Under the r-map, the surface $\mathcal{H}$ gives rise to a 
new complete projective special K\"ahler manifold, which is mapped to  
a new complete quaternionic K\"ahler manifold of dimension 16 under the
c-map.   Let us check the completeness of $\mathcal{H}$. A straightforward
calculation shows that: 
\bean -\partial^2h|_\mathcal{H} &=& 2(1-x^3) \left( \frac{dx^2}{x^2}
+ \frac{dy^2}{y^2}\right) + 2(1+2x^3)\frac{dxdy}{xy}\\
&\ge& 2\left( 1-x^3 -\frac{|1+2x^3|}{2}\right) \left( \frac{dx^2}{x^2}
+ \frac{dy^2}{y^2}\right)\\ 
&\ge& \frac{dx^2}{x^2}
+ \frac{dy^2}{y^2} .
\eean
So the projective special real metric $g= -\frac{1}{3}\partial^2h|_\mathcal{H}$
is bounded from below by the product metric 
\[ \frac{dx^2}{3x^2} +  \frac{dy^2}{3y^2}\]
on $\bR^{<0}\times \bR^{>0}$, which is complete. In fact, 
$\frac{dx^2}{3x^2} +  \frac{dy^2}{3y^2} = d\tilde{x}^2 + d\tilde{y}^2$ under
the change of variables $\tilde{x}=\frac{1}{\sqrt{3}}\ln (-x)$, 
$\tilde{y}=\frac{1}{\sqrt{3}}\ln y$, which maps 
$\bR^{<0}\times \bR^{>0}$ to $\bR^2$.
 
\section{Globalisation of the Ferrara-Sabharwal metric \label{lastSec}}
In this section we will investigate the problem of gluing
Ferrara-Sabharwal manifolds $(N_\a = M_\a \times G, g_\a=g_{M_\a} + g^\a_G)$ 
obtained from projective special K\"ahler 
domains $M_\a \subset M$ in a projective special K\"ahler manifold $(M,g_M)$ to a global quaternionic K\"ahler manifold $(N,g_N)$. Here $g_{M_\a} = g_M|_{M_\a}$.
Recall that $G=\bR^{2n+4}$ with the group structure
defined in section \ref{ccomplsubsec}. 
Denote by $\tilde{M}_\a = \pi^{-1}_M(M_\a )\subset \tilde{M}$  the corresponding 
special coordinate domain in the underlying conical special K\"ahler manifold
$\pi_M : \tilde{M}\ra M$. The affine special coordinates on 
$\tilde{M}_\a$  will be denoted by $q^a$, or, more precisely, by $q^a_\a$. 
The holomorphic special coordinates will be $z^I$ or $z^I_\a$. 
Let $(M_\a)_\a$ be a covering of $M$ by projective special K\"ahler 
domains. Then we define the quotient  
\[ N = \bigcup_\a N_\a/\sim\]
by the equivalence relation
\[  N_{\alpha} \ni (m,v) \sim (m',v') \in N_{\beta} :\Longleftrightarrow m=m'\quad\mbox{and}\quad
v=\tilde{A}_{\a \b}v',\] 
where  $A_{\a \b}$ is the linear symplectic transformation such that 
$q_\a = A_{\a \b}q_\b$ and $\tilde{A}=\mathrm{diag}((A^T)^{-1},\id_2)$.\\  

\bt \label{qKThm} 
The natural projection $\pi : N\ra M$ is a symplectic vector bundle 
and at the same time a bundle of Lie groups. Each fiber is
isomorphic to the solvable Lie group $G$. There exists a 
unique quaternionic K\"ahler structure $(Q,g_N)$ on $N$ such that
$g_N|_{N_\a} = g_\a$. Up to an isomorphism of quaternionic K\"ahler
manifolds consistent with the bundle structures, the quaternionic K\"ahler 
manifold $(N, Q, g_N)$ does neither depend on the covering $(M_\a)_\a$ of $M$ 
nor on the choice of special coordinates on the domains $M_\a$.   
\et 
\pf 
The transition functions $(\tilde{A}_{\a \b})$ can be considered as  
a \v{C}ech 1-cocycle with values 
in the group 
\[ \Sp (2n+2,\bR)\hookrightarrow \Sp(2n+4,\bR), \]
which defines the structure of a symplectic vector bundle on $N$. 
The above linear action of $\Sp(2n+2,\bR)$ on $G=\bR^{2n+4}$ 
is by automorphisms of the solvable Lie group $G$, which means
that the gluing preserves the group structure of the fibers. 
In order to prove that the local metrics $g_\a$ can be glued
to a global Riemannian metric $g_N$ it suffices to 
check that $g_G^\b=\tilde{A}^*g_G^\a$, since we know
already that $g_{M_\a}=g_{M_\b}$ on $M_\a\cap M_\b$.  It is 
useful to rewrite $g_G=g_G^\a$ in the following way:
\begin{equation}
\label{gGreal}
g_G = \frac{1}{4\phi^2} d\phi^2 + \frac{1}{4\phi^2} ( d \tilde{\phi}
+ \sum p_a \Omega^{ab} d p_b )^2
+ \frac{1}{2\phi} \sum \hat{H}^{ab} dp_a d p_b  \;.
\end{equation}
Here $(p_a) = (\tilde{\zeta}_I, \zeta^J)$, 
\begin{eqnarray}
\Omega^{-1} &=& (\Omega^{ab}) := 
\left( \begin{array}{cc}
0 & -\id_{n+1} \\
\id_{n+1} & 0 \\
\end{array} \right) \; ,\\ \nonumber 
\label{DecompHhatInv}
\hat{H}^{-1} &=& 
(\hat{H}^{ab}) := 
\left( \begin{array}{cc}
\mathcal{I}^{-1} &  \mathcal{I}^{-1} \mathcal{R} \\
 \mathcal{R} \mathcal{I}^{-1} &  \mathcal{I} + \mathcal{R} \mathcal{I}^{-1} 
\mathcal{R}) \\ 
\end{array} \nonumber 
\right) \; ,
\end{eqnarray}
that is 
\begin{equation}
\label{DecompHhat}
\hat{H}= (\hat{H}_{ab}) = \left( \begin{array}{cc}
\mathcal{I} + \mathcal{R} \mathcal{I}^{-1} \mathcal{R} &
- \mathcal{R} \mathcal{I}^{-1} \\
- \mathcal{I}^{-1} \mathcal{R} & \mathcal{I}^{-1} \\
\end{array} \right)\; .
\end{equation}
We observe that $2\O$ is the matrix representing the K\"ahler form 
$\o = 2\sum dx^I\wedge dy_I$ of the conical special 
K\"ahler domain $\tilde{M}_\a$ 
in the affine special coordinates $q^a=(x^I,y_J)$. 
The first two terms of \re{gGreal} are manifestly invariant
under symplectic transformations $A\in \Sp (2n+2,\bR)$ since
the 1-form $\sum p_a\O^{ab}dp_b$ is invariant. The invariance 
of the last term is stated in the next lemma establishing
the existence of the metric $g_N$. The proof of the lemma
will be given in the next section together with a
geometric interpretation. 
\bl 
\label{L4}
The tensors $\hat{H}_\a^{-1}$ and $\hat{H}_\b^{-1}$ defined on 
$M_\alpha$ and $M_\beta$ are related by 
\[\hat{H}_\a^{-1} =  A_{\a \b}\hat{H}_\b^{-1} A_{\a\b}^T
\]
on overlaps $M_\alpha \cap M_\beta$. 
\el
The metric $g_N$ is locally a quaternionic K\"ahler metric. 
To see that the local quaternionic structures are consistent, 
we observe that the coordinate transformations $N_\b \ra N_\a$ 
are orientation preserving isometries and that 
an orientation preserving isometry between
two quaternionic K\"ahler manifolds of nonzero scalar
curvature  automatically maps the quaternionic structures
to each other. This follows from the fact that
the restricted holonomy group of a quaternionic 
K\"ahler manifold of nonzero scalar
curvature together with the orientation uniquely 
determines the quaternionic structure. (Notice that
the orientation is needed, since the symmetric quaternionic K\"ahler manifold
\[ \frac{\SO_0 (4,n)}{\SO (4) \times \SO (n)}\]
admits precisely two parallel skew-symmetric quaternionic
structures, which, however, induce opposite orientations.) 
\qed 

The correspondence $(M,g_M)\mapsto (N,g_N)$ established in Theorem 
\ref{qKThm} is a global version of the c-map of Ferrara and Sabharwal. 
We will still call it the {\cmssl supergravity c-map}. 

\bt \label{globalThm}  The supergravity c-map 
maps (isomorphism classes of) 
complete projective special  K\"ahler manifolds $(M,g_M)$ of dimension
$2n$ to (isomorphism classes of) 
complete quaternionic K\"ahler manifolds $(N,g_N)$ of dimension $4n+4$ 
of negative scalar curvature such that $N$ is a vector bundle over $M$ 
with totally geodesic zero section isometric to $M$. 
\et 

\pf This follows from Theorem 
\ref{qKThm} and Theorem \ref{auxThm}. \qed 

\subsection{From Griffiths to Weil flags in special K\"ahler
geometry} 
Let us consider the complex vector space $V=\bC^{2n+2}=\bR^{2n+2}\otimes \bC$ 
with its standard symplectic structure $\O =\sum dz^I\wedge dw_I$ and 
pseudo-Hermitian sesquilinear metric 
\[ \g (u,v) = \sqrt{-1}\O (u,\bar{v}),\quad u,v\in V,\]
of split signature.  We denote by $Gr_0^{k,l}(V)$ the Grassmannian of 
complex Lagrangian subspaces of signature $(k,l)$, where $k+l=n+1$. 
For $k\ge 1$, let $F_0^{k,l}(V)$ denote  
the complex manifold of flags $(\ell ,L)$, where 
$L\in Gr_0^{k,l}(V)$ and  $\ell \subset L$ is a positive definite line.  
Notice that we have a canonical holomorphic projection
\[ F_0^{k,l}(V)=\frac{\Sp (\bR^{2n+2})}{\U (1) \times \U (k-1,l)} 
\lra Gr_0^{k,l}(V)=\frac{\Sp (\bR^{2n+2})}{\U (k,l)} ,\quad (\ell , L) \mapsto 
L,\]
which is $\Sp (\bR^{2n+2})$-equivariant. 

\bp There exists a canonical $\Sp (\bR^{2n+2})$-equivariant 
diffeomorphism 
\[ \psi :  F_0^{k,l}(V) \lra F_0^{l+1,k-1}(V).\]
\ep 

\pf For $(\ell , L)\in F_0^{k,l}(V)$ we put
\[ E := \{ v\in L| v \perp \ell \}\]
and define $\psi (\ell , L) := (\ell , L')$ where 
\[ L' := \ell + \bar{E}.\]
\qed 

In particular, we obtain an equivariant diffeomorphism from the
manifold of {\cmssl Griffiths flags} to the manifold
of {\cmssl Weil flags}:
\[  \psi :  F_0^{1,n}(V) \lra F_0^{n+1,0}(V).\]

\noindent 
{\bf Remark:}
In order to motivate the terminology we observe that
given $L\in F_0^{1,n}(V)$ and a lattice $\G \subset \bR^{2n+2}$ 
the quotient of $W=V/L$ by (the image of) $\G$ is a complex
torus which is analogous to the Griffiths intermediate Jacobian
\[ \frac{H^3(X,\bC )}{H^{3,0}(X,\bC ) + H^{2,1}(X,\bC ) + H^3(X,\bZ )}\]  
whereas the quotient of $W'=V/L'$ by $\G$ is analogous to the
Weil intermediate Jacobian 
\[ \frac{H^3(X,\bC )}{H^{3,0}(X,\bC ) + H^{1,2}(X,\bC ) + H^3(X,\bZ )}\]  
associated to the Hodge structure
of a Calabi-Yau 3-fold $X$. It is known that the bundle
of Griffiths intermediate Jacobians over the (conical special K\"ahler) 
deformation space $M_X=\{ (J,\nu )\}$
of complex structures $J$ of $X$ gauged by a $J$-holomorphic volume form $\nu$ 
carries a hyper-K\"ahler metric obtained from the affine version of the c-map 
\cite{C2}. Similarly, a certain bundle of Weil intermediate Jacobians
is quaternionic K\"ahler by the supergravity c-map \cite{H,V}.\\

Recall 
that there is a totally geodesic $\Sp (\bR^{2n+2})$-equivariant embedding
\be \label{iotaEqu} \iota : Gr_0^{k,l}(V) = \frac{\Sp (\bR^{2n+2})}{\U (k,l)} 
\lra \mathrm{Sym}^1_{2k,2l}(\bR^{2n+2})
= \frac{\SL (2n+2,\bR )}{\SO (2k,2l)}\ee
into the space $\mathrm{Sym}^1_{2k,2l}(\bR^{2n+2})$ of symmetric 
unimodular matrices of signature $(2k,2l)$. The embedding $\iota$ is described
geometrically in \cite{CS}. In the next lemma we give an explicit description 
of $\iota$ in terms of coordinates. Local holomorphic coordinates 
near a point $L_0\in Gr_0^{k,l}(V)$ can be described as follows. Since
$Gr_0^{k,l}(V)$ is a homogeneous space we can assume that 
\[ L_0=  \mathrm{span}\left\{ \frac{\partial}{\partial z^0} + 
i\frac{\partial}{\partial w_0},\ldots , \frac{\partial}{\partial z^{k-1}} + 
i\frac{\partial}{\partial w_{k-1}}, \frac{\partial}{\partial z^{k}} -
i\frac{\partial}{\partial w_{k}},\ldots ,\frac{\partial}{\partial z^{n}} -
i\frac{\partial}{\partial w_{n}}\right\} 
 .\] 
An open neighbourhood $U$ of $L_0$ is given
by
\[ U := \{ L\in Gr_0^{k,l}(V)| L\cap (\bC^{n+1})^* =0\},\]
where 
\[(\bC^{n+1})^* =\{ (z,w)\in V| z=0\}\subset V= T^*\bC^{n+1}=
\bC^{n+1} \oplus  
(\bC^{n+1})^*.\] 
We will now explain that any point 
$L\in U$ is described by a complex symmetric matrix $S= (S_{IJ})$ 
such that the real matrix $\mathrm{Im} S_{IJ}$ has signature $(k,l)$. 
Let us denote by $\mathrm{Sym}_{k,l}(\bC^{n+1})$ the complex vector 
space of all such matrices. Any $S\in \mathrm{Sym}_{k,l}(\bC^{n+1})$ 
defines a Lagrangian subspace 
\[ L = L(S)= \{ (z,w)\in V| w_I = \sum S_{IJ}z^J\}\subset V\]
and one can easily check that the map $S\mapsto L(S)$ is a biholomorphism
$\mathrm{Sym}_{k,l}(\bC^{n+1})\ra U \subset Gr_0^{k,l}(V)$.
Notice that $L_0 = L(S_0)$, $S_0 = iI_{k,l}=i\mathrm{diag}(\id_k,-\id_l)$.  
It is well know that the matrix $S$ transforms as
\be S \mapsto S'= (C+DS)(A+BS)^{-1}\label{S'Equ} \ee 
under a symplectic transformation 
\be  \label{Oequ} {\cal O}= \left( \begin{array}{cc}
A & B \\
C & D \\
\end{array}
\right) \;. \ee
\bl 
\label{L5}
The restriction of the map \re{iotaEqu} to the 
the open subset $U \subset Gr_0^{k,l}(V)$ 
is given by 
\begin{eqnarray} \iota|_U : U\cong \mathrm{Sym}_{k,l}(\bC^{n+1}) &\ra& 
\mathrm{Sym}_{2k,2l}^1(\bR^{2n+2}),\nonumber \\
S = \mathcal{R} + i \mathcal{I} &\mapsto&  \iota (L(S)) = g^S = (g_{ab}^S) 
:=  \left( \begin{array}{cc}
\mathcal{I} + \mathcal{R} \mathcal{I}^{-1} \mathcal{R} &
- \mathcal{R} \mathcal{I}^{-1} \\
- \mathcal{I}^{-1} \mathcal{R} & \mathcal{I}^{-1} \label{gsEqu} \\
\end{array} \right)
.
\end{eqnarray}
\el 

\pf The above formula shows that $g^{S_0}=\mathrm{diag}(I_{k,l},I_{k,l})$.
To prove that $g^S=\iota (L(S))$ for all $S\in U$ it suffices
to check that 
\[g^{S'}={\cal O}^{T,-1}g^S{\cal O}^{-1}\]
for $S'$ defined in \re{S'Equ}.  This follows from Lemma \ref{L6} which
is stated and proved below.
\qed   

The relation between the manifolds of Griffiths and Weil flags, 
the associated Grassmannians, and spaces of symmetric matrices is summarized in the following
diagram. 
\be \label{diagram} 
\xymatrix{
F^{1,n}_0(V) \ar[r]^\psi \ar[d] & F^{n+1,0}_0(V) \ar[d] \\
Gr_0^{1,n}(V) \ar@{^{(}->}[d]^\iota
& Gr_0^{n+1,0}(V) \ar@{^{(}->}[d]^\iota \\
Sym^1_{2,2n}(\bR^{2n+2})
& 
Sym^1_{2n+2,0}(\bR^{2n+2}) \\
}
\ee

\bl
\label{L6}
Let $\mathcal{N}=\mathcal{R}+i\mathcal{I}$ be a complex symmetric
$(n+1)\times (n+1)$ matrix with invertible imaginary part. Using the
decomposition into real and imaginary part, define the real symmetric
$(2n+2)\times (2n+2)$ matrix 
\[
\hat{H} = (\hat{H}_{ab}) = \left(
\begin{array}{cc}
\mathcal{I} + \mathcal{R} \mathcal{I}^{-1} \mathcal{R} & - \mathcal{R}
\mathcal{I}^{-1} \\
- \mathcal{I}^{-1} \mathcal{R} & \mathcal{I}^{-1} \\
\end{array} \right) \;.
\]
Then $\hat{H}$ is invertible with inverse matrix
\be \label{inverseEqu} 
\hat{H}^{-1} = (\hat{H}^{ab}) = \left(
\begin{array}{cc} 
\mathcal{I}^{-1} & \mathcal{I}^{-1} \mathcal{R} \\
\mathcal{R} \mathcal{I}^{-1} & \mathcal{I} + \mathcal{R} \mathcal{I}^{-1}
\mathcal{R} \\
\end{array} 
\right) \;.
\ee
Moreover, $\mathcal{N}$ transforms fractionally linearly under symplectic
transformations \re{Oequ},
\[
\mathcal{N} \rightarrow (C+D \mathcal{N}) (A + B \mathcal{N})^{-1}
\]
if and only if $\hat{H}$, and, hence $\hat{H}^{-1}$ transform as 
tensors:
\[
\hat{H} \rightarrow \mathcal{O}^{T,-1} \hat{H} \mathcal{O}^{-1} \;,\;\;\;
\hat{H}^{-1}  \rightarrow \mathcal{O} \hat{H}^{-1} \mathcal{O}^T \;.
\]
\el
{\bf Remarks:} 
This lemma relates the transformation 
properties of vector multiplet couplings in special holomorphic
and special real coordinates. While we need the lemma to establish the
well-definiteness of the local c-map, it applies to the rigid c-map
as well. In the rigid case the role of $\mathcal{N}$ is played by 
two times the matrix of second derivatives of the holomorphic prepotential, 
$2 (F_{IJ})$, while the role of $\hat{H}$ is played by the Hessian 
metric $\partial^2 H$, where $H$ is the Legendre transform of two times
the imaginary part of the holomorphic prepotential.  
When passing to supergravity, $2 (F_{IJ})$ is replaced by
$\bar{\mathcal{N}}$, while the Hessian metric is replaced by $\hat{H}$. 
When using the superconformal calculus to construct vector multiplet
couplings these replacements are induced by eliminating certain auxiliary
fields. From a geometrical perspective these replacements can be understood
as follows. The kinetic terms of both scalar fields and vector fields
must be positive definite in a physically acceptable theory. In a theory
of rigid supermultiplets, the relevant coupling matrix for both types of 
fields is the metric $2\mbox{Im}F_{IJ}$
of the affine special K\"ahler manifold, 
which therefore must be positive definite. In the 
locally supersymmetric theory scalar and vector fields have different
couplings matrices. The coupling matrix for the scalars is the metric
$\bar{g}$ of the projective special K\"ahler manifold, while 
the coupling matrix for the vector fields is
$\mathcal{N}$, with the kinetic terms given by the imaginary part
$\mathcal{I}$. Therefore $\bar{g}$ and $\mathcal{I}$ must be 
positive definite, which is equivalent to imposing that the 
corresponding conical affine special K\"ahler metric has 
complex Lorentz signature. 

\pf
We now prove Lemma \ref{L6}. It is trivial to verify that \re{inverseEqu}
is the inverse matrix of $\hat{H}$. 
Note that $\mathcal{I}$ is invertible by assumption. The relation between
the transformation properties of $
\mathcal{N}$ and $\hat{H}$ can be 
verified by direct calculation. Such calculations have occured in the
supergravity literature, see for example \cite{DDKV}, 
so that we only need to indicate the main steps. Let $\mathcal{N}'
= \mathcal{R}' + i \mathcal{I}'$ be the matrix obtained by fractionally
linear action of the symplectic matrix $\mathcal{O}$  
on $\mathcal{N}$. This is equivalent
to $\hat{H}$ transforming as a tensor if and only if the following three
relations hold
\begin{eqnarray}
(\mathcal{I} + \mathcal{R} \mathcal{I}^{-1} \mathcal{R})' &=&
D (\mathcal{I} + \mathcal{R} \mathcal{I}^{-1} \mathcal{R}) D^T 
+ D \mathcal{R} \mathcal{I}^{-1} C^T + C \mathcal{I}^{-1} R D^T
+ C \mathcal{I}^{-1} C^T \;,\label{Rel1} \\
(\mathcal{R}\mathcal{I}^{-1})' &=& 
D  (\mathcal{I} + \mathcal{R} \mathcal{I}^{-1} \mathcal{R}) B^T
+ D \mathcal{R} \mathcal{I}^{-1} A^T 
+ C \mathcal{I}^{-1} \mathcal{R} B^T 
+ C \mathcal{I}^{-1} A^T \;,\label{Rel2} \\
(\mathcal{I}^{-1})' &=& 
B (\mathcal{I} + \mathcal{R} \mathcal{I}^{-1} \mathcal{R}) B^T
+ B \mathcal{R} \mathcal{I}^{-1} A^T
+ A \mathcal{I}^{-1} \mathcal{R} B^T
+ A \mathcal{I}^{-1} B^T \;.\label{Rel3} 
\end{eqnarray}
The matrices $A,B,C,D$ are block sub-matrices of the symplectic matrix
$\mathcal{O}$, which satisfies 
\[
\mathcal{O}^T \Omega \mathcal{O} = \Omega \;,\;\;\;
\mbox{where}\;\;\;\Omega =\left( \begin{array}{cc}
0 & \mathbbm{1}_{n+1} \\
- \mathbbm{1}_{n+1} & 0 \\
\end{array} \right) \;.
\]
Therefore they satisfy
\[
A^T C = C^T A \;,\;\;\;
B^T D = D^T B \;,\;\;\;
A^T D - C^T B = \mathbbm{1} \;.
\]
This can be used to verify the following the useful identities
\begin{equation}
U^T (C+D\mathcal{N}) = (C+D\mathcal{N})^T U \;,\;\;
U^T (C+D \bar{\mathcal{N}}) = 
(C+D\mathcal{N})^T \bar{U} - 2i \mathcal{I} \;,
\label{Identity2}
\end{equation}
where $U=U(\mathcal{N}):=A+B\mathcal{N}$. Two further identities are
obtained by complex conjugation. By repeated use of these identities,
we can show that
\begin{equation}
\label{Identity1}
2i\mathcal{I} = \mathcal{N} - \bar{\mathcal{N}} = 
U^T (C + D \mathcal{N}) U^{-1} \bar{U} 
-U^T (C + D \bar{\mathcal{N}}) \;.
\end{equation}
The imaginary part $\mathcal{I}$ of $\mathcal{N}$ transforms 
under symplectic transformations into 
\[
\mathcal{I}' = - \frac{i}{2} [
(C+D\mathcal{N})U^{-1} - (C+D\bar{\mathcal{N}}) \bar{U}^{-1} ] 
\;.
\]
Using identity \re{Identity1} this can be rewritten as
\[
\mathcal{I}' = U^{-1,T} \mathcal{I} \bar{U}^{-1}
= \bar{U}^{-1,T} \mathcal{I} U^{-1} \;,
\]
where the second equation holds because $\mathcal{I}$ is real. Since 
$\mathcal{I}$ is invertible by assumption, we conclude
\begin{equation}
\label{Iinverse}
(\mathcal{I}^{-1})' =
\bar{U} \mathcal{I}^{-1} U^T = U \mathcal{I}^{-1} \bar{U}^T \;.
\end{equation}
Writing out $U=A+B\mathcal{N}$ and $\mathcal{N} = \mathcal{R} + i \mathcal{I}$
we obtain \re{Rel3}.
Next, we note that the real part $\mathcal{R}$ of $\mathcal{N}$
transforms into
\[
\mathcal{R}' = \frac{1}{2}[ 
(C+D\mathcal{N}) U^{-1} + (C + D \bar{\mathcal{N}}) \bar{U}^{-1} ] \;.
\]
Combining this with \re{Iinverse} we obtain
\[
(\mathcal{R} \mathcal{I}^{-1})' =
\frac{1}{2} [
(C+D\mathcal{N}) \mathcal{I}^{-1} \bar{U}^T +
(C+D\bar{\mathcal{N}})^{-1} U^T] \;.
\]
Expressing $U,\mathcal{N}$ in terms of $A,B,C,D$ and $R,\mathcal{I}$, 
we obtain \re{Rel2}. Finally,
we multiply $(\mathcal{R}\mathcal{I}^{-1})'$ by $\mathcal{R}'$ from the right
and add $\mathcal{I}'$. After repeated use of the identities 
\re{Identity2} we finally obtain \re{Rel3}. 
\qed

Let $L\subset V=\bC^{2n+2}$ be a Lagrangian subspace
defined by $S \in \mathrm{Sym}_{1,n}(\bC^{n+1})$ and 
$\ell = \bC (z,w)^T\subset L$. Then $w=Sz$, $(\ell , L)\in F^{1,n}_0(V)$ 
and $(\ell , L') = \psi (\ell , L) \in F_0^{n+1,0}(V)$, cf.\ \re{diagram}. 
\bl
The positive definite Lagrangian subspace $L'\subset V$ corresponds to
the following matrix $S'\in \mathrm{Sym}_{n+1,0}(\bC^{n+1})$: 
\be S'_{IJ} := \bar{S}_{IJ}  + i \frac{\sum_K N_{IK}z^K\sum_L 
N_{JL}z^L}{\sum_{IJ}N_{IJ}z^Iz^J} ,\quad 
N_{IJ} := 2 \mathrm{Im}\, S_{IJ} .\ee
\el 

\pf The following calculation shows that $\ell =\bC (z,w)^T$ 
is contained in the 
Lagrangian subspace $L(S')$ defined by $S'$: 
\[ S'z=  \bar{S}z + 2i (\mathrm{Im}\, S)z = Sz.\]
Next we consider the orthogonal complement $E$ of the line $\ell$ in $L$. 
A vector $(u,Su)^T
\in L$ belongs to $E$ if and only if $\sum N_{IJ}z^I\bar{u}^J=0$.
In that case we obtain
\[ S'\bar{u}=\bar{S}\bar{u} = \ol{Su},\]
which proves that $\bar{E}$ is contained in $L(S')$. Therefore, $L(S')=L'$. 
\qed 

\bc \label{Cor5} \begin{itemize}
\item[(i)] The complex 
matrix $S'={\cal N}\in \mathrm{Sym}_{n+1,0}(\bC^{n+1})$ occurring in the 
definition of the Ferrara-Sabharwal metric, see \re{FRI}, 
is related to the matrix $S=F_{IJ}\in \mathrm{Sym}_{1,n}(\bC^{n+1})$
by the correspondence $S\mapsto S'$ of the previous lemma, which 
is induced by the map $\psi$ from Griffiths flags to Weil flags, cf.\ 
\re{diagram2}. 
\item[(ii)] The real matrix \re{DecompHhat} occurring in the 
formula \re{gGreal} for the Ferrara-Sabharwal metric is given by
\[ \hat{H}= g^{S'}= g^{{\cal N}}, \]
where the map $S\mapsto g^S$ is defined in \re{gsEqu}.   
\end{itemize}
\ec
 
The following diagram gives an overview of the relations between 
the Lagrangian subspaces $L=L(S)$, $L'=L(S')$ and the corresponding
real matrices $g^S=(g_{ab})$ and $g^{S'}=\hat{H}_{ab}$: 
\be \label{diagram2} 
\xymatrix{
(\ell , L)\in F^{1,n}_0(V) \ar[r]^\psi \ar[d] & F^{n+1,0}_0(V)\ni (\ell , L') \ar[d] \\
L\in Gr_0^{1,n}(V) \ar@{^{(}->}[d]^\iota
& Gr_0^{n+1,0}(V)\ni L' \ar@{^{(}->}[d]^\iota \\
g^S\in Sym^1_{2,2n}(\bR^{2n+2})
& 
Sym^1_{2n+2,0}(\bR^{2n+2})\ni g^{S'}, \\
}
\ee
where the line $\ell$ is generated by the vector $(z,Sz)=(z,S'z)$. 
\bc \label{Cor6}
The (indefinite) 
affine special K\"ahler  metric $g=2\sum g_{ab}dq^adq^b$ is related
to the positive definite metric $g'=2\sum \hat{H}_{ab}dq^adq^b$ by 
\[ g'|_{\mathcal{D}}=g|_{\mathcal{D}},\quad g'|_{\mathcal{D}^\perp} = -
g|_{\mathcal{D}^\perp},\]
where $\mathcal{D}$ is defined in Definition \ref{conDef} (iv). 
\ec 

\pf This follows from the geometric description of the
map $\iota : L=L(S)\mapsto g^S$ \cite{CS}. 
Recall that in the affine special coordinates 
$q^a=(x^I=\mathrm{Re}\, z^I,y_J=\mathrm{Re}\, w_J)$ the K\"ahler form
is given by $\o = 2\sum dx^i\wedge dy_i$.  
The matrix $g=g^S$ represents
the scalar product $\mathrm{Re}\, \g|_L$ in the Darboux coordinates
$\sqrt{2}q^a$ restricted to $L$. 
We can compare the restrictions of $\g$ to 
$L$ and $L'$ by the following isomorphism of real 
vector spaces $\Psi : L \ra L'$: 
\[  \Psi|_\ell  := \mathrm{Id},\quad  \Psi (v) := \bar{v} \quad
\mbox{for all}\quad v\in E.\]   
We can easily see that $\Psi$ is an isometry on $\ell = L\cap L'$ and 
an antiisometry $E\ra \bar{E}$ on $E$. In fact, 
$\g (\bar{v},\bar{v})=-\g (v,v)$ for all $v\in V$. This 
shows that the metric $\Psi^*\mathrm{Re}\, \g|_{L'}$ is related 
to $g=\mathrm{Re}\, \g|_L$ by changing the sign on the orthogonal 
complement of $\ell$.
Finally, the Gram matrix $g^{S'}=(\hat{H}_{ab})$ of  
$\mathrm{Re}\, \g|_{L'}$ in the coordinates 
$\sqrt{2}q^a|_{L'}$ is the same as that of $\Psi^*\mathrm{Re}\, \g|_{L'}$
in the coordinates $\sqrt{2}q^a|_L$, since $q^a\circ \Psi = q^a$.  
\qed  

We now finally prove Lemma \ref{L4} and thus complete the
proof of Theorem \ref{qKThm}.

\pf
Let us denote by $F^\a$ the holomorphic prepotential of  
the special K\"ahler domain $M_\a$ and put $S^\a := F_{IJ}^\a$. We know that 
the complex matrices 
$S_\a$ and $S_\b$ are related by a fractional linear transformation
associated with a symplectic transformation $\mathcal{O}=
A_{\a \b }\in \Sp (\bR^{2n+2})$, which relates 
the Gram matrices $g^{S_\a}$ and $g^{S_\b}$ of the corresponding affine special 
K\"ahler metrics $g_{\tilde{M}_\a}$ and $g_{\tilde{M}_\b}$ 
in affine Darboux coordinates. Since all the maps 
in the diagram \re{diagram2} are 
$\Sp (\bR^{2n+2})$-equivariant, this implies that the 
matrices ${\cal N}_\a$ and  ${\cal N}_\b$ are related by the same
fractional linear transformation and that 
$g^{S_\a'}$ and $g^{S_\b'}$ are related by the symplectic 
transformation $\mathcal{O}$. 
This shows that $\hat{H}_\a$ transforms as claimed
in Lemma \ref{L4}. 
\qed

\subsubsection*{Acknowledgements}

This work was supported by the German Science 
Foundation (DFG) under the Collaborative Research Center (SFB) 676.
The work of T.M. was supported in part by STFC grant ST/G00062X/1.

\end{document}